\documentstyle[preprint,aps,prl]{revtex} 

\begin{document}

\draft

\title{Nonequilibrium spin distribution in single-electron transistor}

\author{Alexander N. Korotkov\footnote{On leave from Nuclear Physics
Institute, Moscow State University, Moscow 119899, Russia} and 
V. I. Safarov}
\address{
GPEC, Departement de Physique, Facult\'e des Sciences de Luminy, 
Universit\'e de la M\'editerran\'ee, 
13288 Marseille, France} 

\date{\today}

\maketitle

\begin{abstract}
        Single-electron transistor with ferromagnetic outer electrodes
and nonmagnetic island is studied theoretically. Nonequilibrium electron 
spin distribution in the island is caused by tunneling current. 
The dependencies of the magnetoresistance ratio $\delta$ on the bias and 
gate voltages show the dips which are 
directly related to the induced separation of Fermi levels for electrons
with different spins. Inside a dip $\delta$ can become negative.
\end{abstract}


\narrowtext

        Magnetoresistance of tunnel structures is currently an attractive 
topic for both experimental and theoretical studies (see, e.g., Refs.\  
\cite{Julliere,Moodera,Helman,Slonczewski,Milner}). 
For tunnel junctions made of ferromagnetic films, 
the difference as high as 26\% at 4.2 K (up to 18\% at
room temperature) between the tunnel resistances for parallel
and antiparallel film magnetization has been observed \cite{Moodera},
that allows their application as magnetic sensors. 
The low temperature values agree well
with the theoretical result \cite{Julliere} 
$\Delta R/R = 2P_1P_2 /(1+P_1P_2)$ where $P_1$ and $P_2$ are spin
polarizations of tunneling electrons in two films, that proves 
the good achieved quality of junctions. 
As an example, the
polarization is about 47\% for CoFe, 40\% for Fe, and 34\% for Co
\cite{Meservey,Moodera}.

	With the decrease of the tunnel junction area, the single-electron
charging \cite{Av-Likh} becomes important leading to new physical
effects. The study of tunnel magnetoresistance in this regime is
a rapidly growing field 
\cite{Ono,Ono-c/m,Schelp,Sankar,Barnas,Takahashi,Majumdar} 
(see also Refs.\ \cite{Helman} and \cite{Milner}).
	For example, the enhancement of the magnetoresistance ratio 
$\Delta R/R$ due to Coulomb blockade has been discussed in Refs.\ 
\cite{Ono,Takahashi}.
        The magnetic field dependence of the tunneling current between two 
Co electrodes via the layer of few-nm Co clusters has been measured 
in Ref.\ \cite{Schelp}.
        In Ref.\ \cite{Barnas} the model of small ferromagnetic double tunnel
junction (i.e. single-electron transistor \cite{Likh-87} (SET-transistor)
without gate electrode) has been considered, 
in which the tunnel resistances of junctions are different 
for parallel and antiparallel magnetizations of electrodes, thus 
changing the current through the system. A similar SET-transistor has
been also studied theoretically in Ref.\ \cite{Majumdar}.
	The very interesting effect of magneto-Coulomb oscillations 
in SET-transistor has been observed and explained in Refs.\ 
\cite{Ono,Ono-c/m}. Strong magnetic field $H$ causes the Zeeman shift   
of two spin bands and their repopulation. Since 
in ferromagnets the densities of states of these bands are different, 
the Fermi level moves with $H$ leading to magneto-Coulomb oscillations.

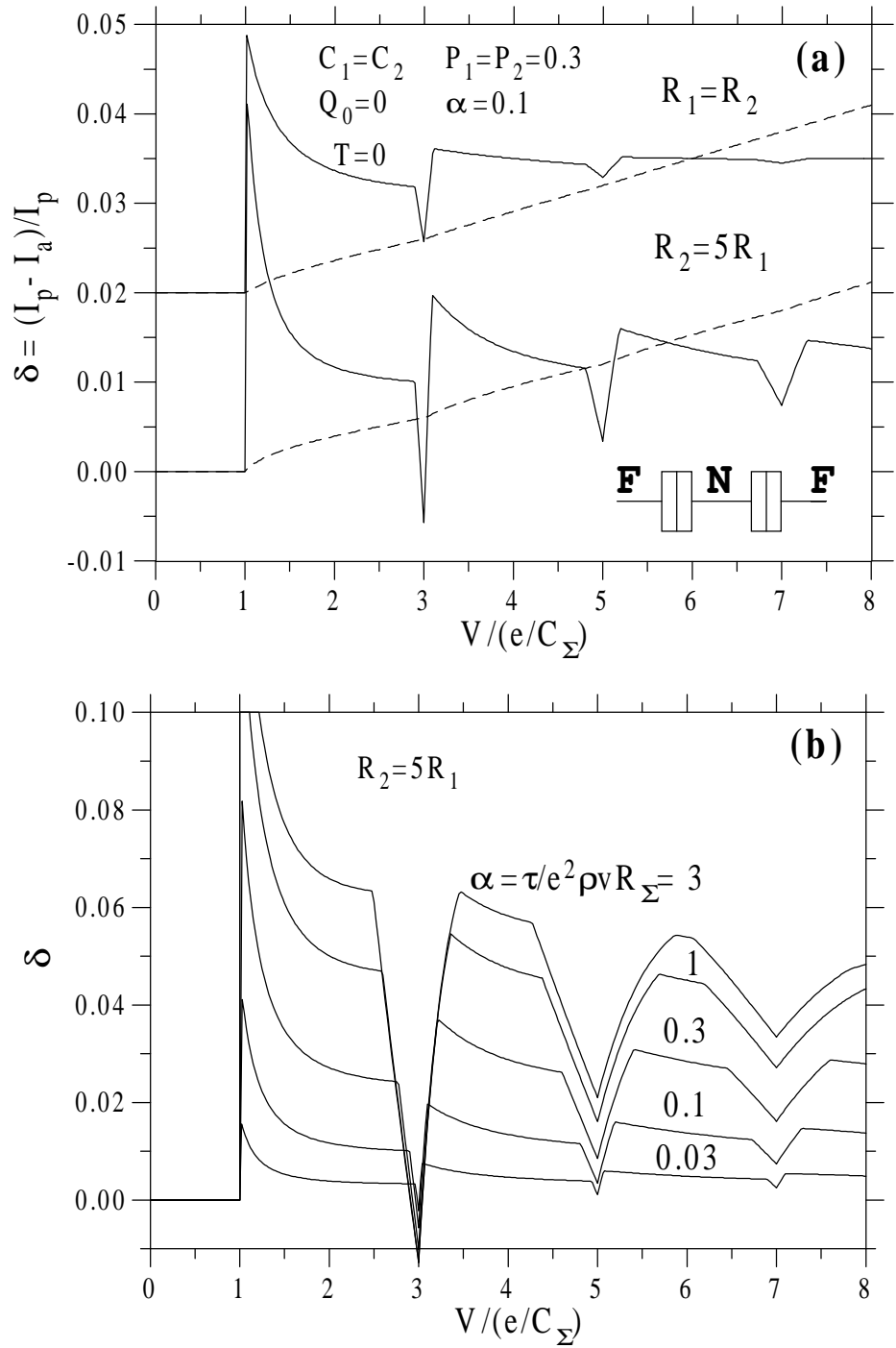
\begin{figure}
\caption{ The dependence of the magnetoresistance ratio $\delta$ on
the bias voltage $V$ for (a) two different ratios $R_2/R_1$ and
(b) for several values of the dimensionless spin relaxation time $\alpha$. 
Inset in (a) shows the schematic of the SET-transistor with ferromagnetic (F)
outer electrodes and nonmagnetic (N) island, while the dashed lines 
show the $I_p$--$V$ curves (arbitrary units).}  
\label{fig1}\end{figure}

        In the present letter we consider a
SET-transistor which has ferromagnetic outer 
electrodes and nonmagnetic central island (see inset in Fig.\ 
\ref{fig1}a). 
When the coercive fields of two ferromagnetic electrodes are different, 
the standard technique of the magnetic field sweeping (see, e.g., Ref.\ 
\cite{Moodera}) easily allows to obtain parallel or antiparallel 
polarizations of outer electrodes.
In the first approximation the current through SET-transistor 
does not depend on these polarizations because the island
is nonmagnetic (the Zeeman splitting is negligible because 
of small $H$). However, if the electron spin relaxation in the island
is not too fast (estimates are discussed later), then the tunneling 
of electrons with preferable spin orientation
creates the nonequilibrium spin-polarized state of the island (similar
to the effect discussed in Refs.\ \cite{Johnson-1,Johnson-2} in absence
of the Coulomb blockade). 
This in turn affects the tunneling in each junction and leads to
different currents $I_p$ and $I_a$ through SET-transistor in 
the parallel and antiparallel configurations. 

        We will calculate the dependence of the 
relative current change $\delta =(I_p-I_a)/I_p$  
on the bias and gate voltages (we call $\delta$ magnetoresistance 
ratio despite for finite voltage this terminology could be misleading). 
Nonzero $\delta$ is already the evidence of the nonequilibrium spin 
state in the island. Moreover, the voltage dependence of $\delta$
shows the dips, the width of which directly corresponds to the energy 
separation between Fermi levels of electrons with different 
spins in the island.

        We consider SET-transistor consisting of two tunnel junctions
with capacitances $C_1$ and $C_2$. Induced background charge $Q_0$
as usual \cite{Likh-87} describes the influence of the gate voltage
(general equivalence relations for finite gate capacitance are 
discussed, e.g., in Ref.\ \cite{Kor-gate}). We assume that the
voltage scale related to the magnetic polarization of ferromagnetic
electrodes \cite{Slonczewski} and the voltage scale of the barrier 
suppression \cite{Kor-Naz} are large in
comparison with the single-electron charging energy (that is
a typical experimental situation). Then the polarization of outer
electrodes can be taken into account by the difference between the tunnel 
resistances $R^u_{1,2}$ and $R^d_{1,2}$ for electrons with ``up'' and
``down'' spins.  The total junction resistances 
$R_1=(1/R_1^u+1/R_1^d)^{-1}$ and $R_2=(1/R_2^u+1/R_2^d)^{-1}$ do not
depend on the magnetic polarizations $P_1$ and $P_2$ of electrodes, while
``partial'' resistances are given by $R_i^u=2R_i/(1+P_i)$ and
$R_i^d=2R_i/(1-P_i)$ (similar to the model of Ref.\ \cite{Julliere}). 
Notice that $P_i$ describes the polarization  
of tunneling electrons \cite{Meservey} which is different 
(typically even in sign) 
from the total electron polarization at Fermi level 
(the latter one determines, e.g., the period of 
magneto-Coulomb oscillations  \cite{Ono-c/m}). 

        We assume that the energy relaxation of electrons in
the island is much faster than the spin relaxation. So, we  
characterize the nonequilibrium spin state by the difference
$\Delta E_F$ between Fermi levels for ``up'' and ``down'' spins  
while both distributions are determined by the thermostat temperature $T$.
The spin diffusion length is assumed to be much larger than the island
size (that is a typical experimental situation -- see Ref.\ 
\cite{Johnson-1}), so the spin distribution is uniform along the island.

        The equations of the ``orthodox'' theory for single-electron
transistor \cite{Av-Likh,Likh-87} (we assume $R_i \gg R_K=h/e^2$) 
should be modified in our case.
         The energy gain $W_{i}^{u(d)\pm}$ for tunneling to (+) or 
from (-) the island through $i$th junction is different for ``up''
and ``down'' electrons,
        \begin{eqnarray}
W_i^{u \pm} (n) = W_i^{\pm}  \mp \frac{\Delta E_F}{2}, \,\,
W_i^{d \pm} (n) = W_i^{\pm}  \pm \frac{\Delta E_F}{2}
	\label{W}   \\
W_i^{\pm} =\frac{e}{C_\Sigma} \left[ \mp (ne+Q_0) \mp (-1)^i 
\frac{VC_1C_2}{C_i} -\frac{e}{2} \right] . \nonumber
        \end{eqnarray}
        Here $n$ is the number of extra electrons on the island 
(as usual the electron is assumed to have positive charge $e$),
$C_\Sigma =C_1+C_2$, and $V$ is the bias voltage.
        The corresponding tunneling rates satisfy usual equation
\cite{Av-Likh}
        \begin{equation}
\Gamma_i^{s \pm} (n) =\frac{W_i^{s \pm}(n)}{e^2 R_i^s 
\left( 1-\exp ( -W_i^{s \pm} (n)/T) \right) } , 
        \label{Gamma}\end{equation}
where $s=u,d$ denotes spin. 
        The average current $I$ through SET-transistor can be calculated as
        \begin{equation}
I=\sum_{n,s} e \left[ \Gamma_1^{s+}(n)-\Gamma_1^{s-} (n) \right] 
        \sigma (n) ,
        \label{Iaver}\end{equation}
where $\sigma (n)$ is the stationary solution of the master equation
\cite{Av-Likh}
        \begin{equation}
d\sigma (n)/dt= \sum_{i,s,\pm} [\sigma(n\pm 1) 
		\Gamma_i^{s\mp} (n\pm 1) 
	- \sigma (n) \Gamma_i^{s\pm} (n)].
        \label{master}\end{equation} 
        Finally, the Fermi level separation $\Delta E_F$ should
satisfy selfconsistent equation
        \begin{eqnarray}
\Delta E_F \rho v/\tau = \sum_{n,i} \left[ \Gamma_i^{u+}(n)-\Gamma_i^{d+} (n)
\right. 
	\nonumber \\ 
\left. -\Gamma_i^{u-} (n) +\Gamma_i^{d+}(n)   \right] 
        \sigma (n)  ,
        \label{dEF}\end{eqnarray}
where $\tau$ is the electron spin relaxation time for the island,
$\rho$ is the density of states (per spin), and $v$ is island's volume.
We introduce also the dimensionless spin relaxation time 
$\alpha = \tau /e^2\rho v(R_1+R_2)$. 

	The signs of polarizations $P_1$ and $P_2$ can be changed 
using the external magnetic field, that interchanges resistances
$R_i^u$ and $R_i^d$. So the current $I_p$ for the parallel magnetization
($P_1 P_2>0$) is different from the current $I_a$ when one magnetization
direction is reversed, $P_2\rightarrow -P_2$ (the change $P_1 
\rightarrow -P_1$ obviously gives the same result). 
	Figure \ref{fig1}a shows the numerically calculated dependence 
of the magnetoresistance ratio $\delta$ (solid line) on the bias voltage 
$V$ for the SET-transistor with
parameters $C_1=C_2$, $Q_0=0$, $T=0$, $|P_1|=|P_2|=30\%$, and $\alpha =0.1$.
For the upper curve (shifted up for clarity) we assumed $R_2=R_1$ while 
$R_2=5R_1$ for the lower curve. The $\delta $--$V$ dependence shows
the oscillations with the same period $e/C_i$ as for the Coulomb 
staircase. The existence of oscillations is a trivial consequence 
of the charge dynamics in SET-transistor, similar to the effect 
discussed in Refs.\ \cite{Barnas,Majumdar}. 

	More interesting features seen in
Fig.\ \ref{fig1}a are the triangular-shape dips near the bias voltages 
	\begin{equation}
V=[e/2+ne+(-1)^i Q_0]/C_i,
	\label{dip-pos}\end{equation}
at which the derivative of the I-V curve 
abruptly increases (the Coulomb staircase for $I_p$ shown in Fig.\ 
\ref{fig1}a by dashed lines 
is better seen for the lower curve). The edges of a dip correspond
to the alignment between the Fermi level in an electrode and one
of the split Fermi levels for electrons with different spins 
in the island. Hence, the dip width $\Delta V$ is directly related
to the Fermi level splitting, $\Delta V = \Delta E_F C_\Sigma /eC_i$.
Somewhat similar effect (the finite voltage caused by injected spin
current) without Coulomb blockade had been observed in the ``spin 
transistor'' \cite{Johnson-1} fabricated using the thin film geometry
with a size scale of 50 $\mu$m; the typical signal scale was only
about 30 pV.
The small size of the SET-transistor island leads to the very strong 
enhancement of the Fermi level separation (see Eq.\ (\ref{dEF})) 
and the corresponding voltage scale. 

	The width of the dips in Fig.\ \ref{fig1}a 
increases with voltage because the larger current 
provides larger $\Delta E_F$ (the crude estimate is $\Delta E_F =
\alpha I(P_1-P_2)eR_\Sigma$). 
Actually in the case $|P_1|=|P_2|$ shown in the figures, 
the nonequilibrium spin distribution 
in the island occurs only for the antiparallel magnetization of electrodes; 
for parallel magnetization $\Delta E_F=0$ because the spin currents 
to/from different electrodes exactly cancel each other.  
When $|P_1|\neq |P_2|$ the complete cancellation does not occur, and  the 
dip shape is determined by two different values of $\Delta E_F$ 
leading to the trapezoid-like shape instead of the triangular one.
 
	It is interesting that the magnetoresistance ratio $\delta$
can be even negative within the dip range (see Fig.\ \ref{fig1}). 
This can be understood in the
following way. The $I_a$--$V$ curve for the antiparallel magnetization 
generally goes below $I_p$--$V$ curve because the Fermi level splitting
(which is larger in the antiparallel case) decreases the effective voltage 
for the spin band which provides the easier
tunneling. However, this also splits the kinks on the I-V curve leading
to the increase of $I_a$ (and decrease of $\delta$) within the splitting 
range. For sufficiently steep kink, $I_a$ can become even larger than $I_p$
(negative $\delta$). This also explains why the dips are more significant 
for larger tunnel resistance ratio (see Fig.\ \ref{fig1}a) when the
Coulomb staircase is more pronounced.

	Increase of the spin relaxation time $\tau$ leads to larger
$\Delta E_F$ and, hence, increases $\delta$ as well as widens the dips, 
that is illustrated in Fig.\ \ref{fig1}b  ($\delta =0$ for $\tau = 0$).
The change of the polarization amplitudes $|P_1|$ and $|P_2|$ leads
to similar effects. Crudely, the magnetoresistance is determined by
the product $\alpha |P_1 P_2|$, while the exact shape of the $\delta$--$V$
curve depends on each of these parameters.

	 In the limit of large bias voltage the magnetoresistance ratio
can be found analytically using the following expression for the current:
	\begin{eqnarray}
&& IR_\Sigma /V = 1- (\alpha /2) (P_1-P_2)^2/
	\nonumber \\
&& \left[ 1+(\alpha /2)
[R_\Sigma^2/R_1R_2 -R_1R_2(P_1/R_1+P_2/R_2)^2] \right] .
	\end{eqnarray}
However, the formula for $\delta$ is rather lengthy, so we present 
here only the result for small $\alpha$, 
	\begin{equation}
\delta =2\alpha |P_1P_2|,
	\label{smalla}\end{equation}
 and the 
expression $\delta = 2\alpha |P_1P_2|/[1+2\alpha (1-(|P_1|-|P_2|)^2/4)]$ 
for the case  $R_1=R_2$.

	The finite temperature smears the features of the 
$\delta$--$V$ dependence (see Fig.\ \ref{fig2}a), but obviously 
does not change $\delta$ in the large-bias limit. 
 The dips disappear when $T$ becomes comparable to $\Delta E_F$ 
while the oscillations
disappear at higher temperatures determined by the single-electron
energy scale $e^2/C_\Sigma$. 

	Notice that two series of dips determined by Eq.\ (\ref{dip-pos}) 
coincide in Figs.\ \ref{fig1} and \ref{fig2}a. With the change of 
the background charge $Q_0$ by the gate voltage, these two series 
will move in opposite directions.   
 	The dips can be also seen on the $\delta$--$Q_0$ dependence 
which is shown in Fig.\ \ref{fig2}b for different
bias voltages $V$. The dip position moves with $V$. 
There are two dips per period, however, one of them
is much less pronounced because of relatively large ratio of tunnel 
resistances.

	To estimate the parameters of a possible experimental
realization, let us assume Co-Cu-Co SET-transistor. (Notice that
Al-Co-Al SET-transistor has been already fabricated \cite{Ono-c/m},
however, aluminum is not suitable for our purpose because of its
superconductivity.) The polarization $|P|=30\%$ used in figures
is a conservative value for Co. The spin relaxation rate $\tau$
for nonmagnetic island, which is the most crucial parameter of the 
effect, depends much on the material quality. In Ref.\ \cite{Lubzens}
$\tau \sim 10^{-7}$ s has been reported for very pure Cu at $T=1.4$ K
(the similar value has been found in Ref.\ \cite{Lubzens} for Al, while
$\tau \sim 10^{-8}$ s have been reported for Al in Ref.\ \cite{Johnson-1}).
Let us choose $\tau =10^{-8}$ s. Then using $\rho = 9\times
10^{21}$ eV$^{-1}$cm$^{-3}$ for Cu, $R_\Sigma =10^5\, \Omega$, 
and the island volume $v=200\mbox{nm}\times
50 \mbox{nm}\times 20\mbox{nm}$, we get $\alpha =0.35$.
Hence, the effect of nonequilibrium spin distribution should be
rather strong, and we could expect the magnetoresistance ratio $\delta$ 
up to $\sim 10\%$ ($\delta$ is significantly enhanced near the Coulomb
blockade threshold -- see Fig.\ \ref{fig1}). This large value 
allows to consider the possible applications of such a device.
	For $C_\Sigma \sim 3\times 10^{-16}$ F the dips of the
$\delta$--$V$ dependence could be observed at temperatures below 
$\sim 0.2$ K while the oscillations
could be noticeable up to $T\sim 1$ K.

	In our theory we have neglected	the Zeeman splitting 
$\pm g\mu_B H/2$ because the typical coercive fields are relatively
small, $H \sim 10^2$ Oe \cite{Moodera}. Hence the 
corresponding energy scale is very small, 
$\Delta E \sim 10^{-6}$ eV $\sim 10^{-2}$ K,
and the effect can hardly be observed. 

	We have discussed dc case only. If ac voltages 
are applied to the bias and/or gate electrodes, the similar formalism
can be used. However, in this case the dynamic solution of the master 
equation (\ref{master}) should be used instead of the stationary solution,  
and also the left side of Eq.\ (\ref{dEF}) should be replaced by 
$[d(\Delta E_F)/dt +\Delta E_F/\tau]\rho v$. The nontrivial dependence 
starts when the frequency $f$ of the applied voltage becomes comparable
to $\tau^{-1}$. These frequencies are within the experimentally 
achievable range, so such an experiment could be used for the 
direct measurement of the spin relaxation time $\tau$.

	In conclusion, we have considered the SET-transistor consisting
of ferromagnetic electrodes and nonmagnetic island. We have predicted
that the nonequilibrium spin distribution in the island leads to a 
considerable magnetoresistance which has a specific dependence on
the bias and gate voltages. In particular, it shows the dips 
directly related to the Fermi level splitting.

The authors thank D. V. Averin and K. K. Likharev for useful discussions.
A. K. has been supported in part by French MENRT (PAST), Russian RFBR, 
and Russian Program on Nanoelectronics.

\begin{figure}
\caption{(a) The $\delta$--$V$ dependence for different temperatures $T$
and (b) the dependence of $\delta$ on the background charge $Q_0$ 
for several bias voltages. }
\label{fig2}\end{figure}

\end{document}